\documentclass[10pt,aps,pra,twocolumn,floatfix,floats,showpacs,superscriptaddress,raggedbottom]{revtex4-1}
\usepackage{graphicx,latexsym}
\usepackage{amssymb,amsmath,bm}
\usepackage{subfigure}
\usepackage{braket}
\usepackage{bbm}

\usepackage{hyperref}
\hypersetup{
    pdfnewwindow=true,       % links in new window
    colorlinks=true,         % false: boxed links; true: colored links
    linkcolor=blue,          % color of internal links
    citecolor=blue,          % color of links to bibliography
    filecolor=magenta,       % color of file links
    urlcolor=black           % color of external links
}

\def\eq#1{Eq.\ (\ref{#1})}

\def\fig#1{Fig.\ \ref{#1}}

\begin{document}

%-----------------------------------------------------------------
\title{Excitation spectra of a quantum ring embedded in a photon cavity}

\author{Thorsten Arnold}
%\email{tla1@hi.is}
\affiliation{Science Institute, University of Iceland, Dunhaga 3,
        IS-107 Reykjavik, Iceland}
\author{Chi-Shung Tang}
%\email{cstang@nuu.edu.tw}
\affiliation{Department of Mechanical Engineering,
        National United University,
        1, Lienda, Miaoli 36003, Taiwan}
\author{Andrei Manolescu}
%\email{manoles@ru.is}
\affiliation{School of Science and Engineering, Reykjavik University,
        Menntavegur 1, IS-101 Reykjavik, Iceland}
\author{Vidar Gudmundsson}
%\email{vidar@hi.is}
\affiliation{Science Institute, University of Iceland, Dunhaga 3,
        IS-107 Reykjavik, Iceland}
%
%----------------------------------------------------------------

\begin{abstract}
We explore the response of a quantum ring system coupled to a photon
cavity with a single mode when
excited by a classical dipole field. We find that the energy
oscillates between the electronic and photonic components of the
system. The contribution of the linear and the quadratic terms in
the vector potential to the electron-photon interaction energy are
of similar magnitude, but opposite signs stressing the importance of retaining both in the
model. Furthermore, we find different Fourier spectra for the
oscillations of the center of charge and the oscillations of the
mean photon number in time. The Fourier spectra are compared to the
spectrum of the many-body states and selection rules discussed. In
case of the center of charge oscillations, the dipole matrix
elements preselect the allowed Bohr frequencies of the transitions,
while for the oscillations of the mean photon number, the difference
of the photon content of the many-body states influences the
selection rules.
\end{abstract}


\maketitle
\section{Introduction}

Charge oscillations in mesoscopic electronic systems have been
investigated for a long time by far-infrared spectroscopy
\cite{PhysRevLett.64.788,RevModPhys.74.1283}.  Even for simple
systems like quantum dots such oscillations are usually complex,
incorporating the interaction of electrons with the confinement
potentials and the many-body (MB)
electron-electron interactions as well. Specific symmetries or
selection rules may sometimes inhibit complex degrees of freedom and
reduce the oscillations to the motion of the center of
mass \cite{PhysRevB.43.12098,PhysRevB.44.13132}. Resonant Raman
scattering has also been used to observe and distinguish
single-electron excitations and collective modes with monopole,
dipole, or quadrupole symmetry characterized by angular momenta
quantum numbers $m=0, \pm 1,\pm
2$ \cite{PhysRevB.59.10240,PhysRevB.61.15600}.

Charge oscillations, or other time-dependent phenomena in a mesoscopic
system, can be induced by a radiation field \cite{Mani2002,Tang2001}.
It was shown theoretically that a short pulse can induce a
charge current~\cite{PhysRevB.67.161301,Zhu2009} or a spin current
\cite{PhysRevB.83.155427,Nita201212} in a ring. Given the inhomogeneous
charge distribution in a nanoscale system, nonlinear response of
the charge or currents to an external electromagnetic pulse is to be
expected \cite{Gudmundsson2014}.

Complementary to charge or current oscillations, the interaction of electrons
with photons has been at the core of condensed matter physics research
for a long time. Electronic systems on the nanoscale offer a unique
opportunity to study this interaction, and the possibility to control
optical response.  For example the dependence of optical transitions
in nanostructures on shape~\cite{Ingibjorg99:16591,Gomis2006} and
temperature~\cite{3935476920090504} has been studied.  In some
cases, quantum rings have been preferred over quantum dots when
selecting a quantum confinement since the oscillator strength for
the excitonic ground state is larger~\cite{Pettersson2000510}.
The photoluminescence and the excitation spectrum of quantum
rings and their optical transitions and selection rules have been
investigated~\cite{PhysRevB.72.155331,1.4789519}. However, it is
found that field effect structure devices reduce the oscillator
strength~\cite{PhysRevB.75.045319}.

The electron-photon interaction may be to some extent controlled inside
optical microcavities, and such experiments on single photon emission
in quantum rings have been carried out \cite{1742-6596-210-1-012037}.
Also, photoluminescence measurements on quantum rings and optical
emission of quantum dots coupled to photonic microcavities have been
performed~\cite{Sarkar20082156,Gallardo:10}.
Electron-photon systems in cavities, which are in fact quantum electrodynamical
systems at mesoscopic scale, constitute an emerging topic of active
research. Either driven by the need of elements for quantum computing
\cite{PhysRevB.75.045331}, or due to the possibility of polariton
condensation at high temperatures \cite{Laussy2012}, or polariton blocking
\cite{PhysRevB.73.193306}, or in the context of photo-assisted transport
through a nanoscale system \cite{Platero2004}.  Recently, it has been
shown that the charge and spin currents in quantum rings can be tuned by
the interaction of electrons with photons carrying linear or circular polarization
\cite{PhysRevB.87.035314,EPJB_87_113,Arnold2014170}.

The emission spectrum of an Aharonov-Bohm~\cite{PhysRev.115.485} quantum
ring in a single mode microcavity was theoretically studied in the
strong coupling regime under the influence of time-independent external
fields~\cite{PhysRevB.88.085429}. The effect of a short light pulse on
the emission of dipolaritons in quantum wells embedded in a microcavity
has been discussed~\cite{PhysRevLett.111.176401}. Oscillations of the
electric dipole moment for a single electron Aharonov-Bohm ring have
been related to the selection rules of the optical transitions in
the ring and have been shown to allow a control of the polarization
properties~\cite{PhysRevB.85.245419}.

To the best of our knowledge, the effect of a short time-dependent
electromagnetic pulse on a quantum ring coupled to a photon cavity
%via both the paramagnetic and the diamagnetic
%part of the electron-photon interaction has not yet been investigated.
has not been investigated. In this paper, we study numerically the
time evolution of the energies of the MB electron-photon system.  We
calculate the dipole moment of the charge density and the mean
photon number beyond two level models and the rotating wave
approximation~\cite{Nakano1998328,1367-2630-14-1-013036}. The time
evolution can be considered as the non-linear response to a
classical excitation pulse. The time evolution is analyzed using
Fast Fourier transform and the selection rules between the MB states
discussed.  Sec.\ II describes the quantum ring model coupled to a
photon cavity and the excitation pulse.  The electron-photon
interaction is described using the method of ``exact''
diagonalization. Sec.\ III presents the results and a summary is
given in Sec.\ IV.

%Kohn theorem does not apply, no simple symmetries as for quantum dot.

\section{Theory and model} \label{sec2}

Here, we describe the Hamiltonian of the MB electron-photon system including the potential used to model
the finite width quantum ring and its time evolution during and after excitation by a short pulse.

\subsection{Many-body system Hamiltonian}

The MB Hamiltonian of the system in second quantization without the time-dependent excitation is
\begin{align}
      \hat{H}_{\rm MB}=&\int d^2 r\; \hat{\mathbf{\Psi}}^{\dagger}(\mathbf{r})\left[\frac{\hat{\mathbf{p}}^2}{2m^{*}}
      +V(\mathbf{r})\right]\hat{\mathbf{\Psi}}(\mathbf{r})\nonumber \\
      &+\hbar\omega \hat{a}^{\dagger}\hat{a}, \label{H^S}
\end{align}
with the spinor
\begin{equation}
      \hat{\mathbf{\Psi}}(\mathbf{r})=
      \left( \begin{array}{c} \hat{\Psi}(\uparrow,\mathbf{r}) \\ \hat{\Psi}(\downarrow,\mathbf{r}) \end{array} \right)
\end{equation}
and
\begin{equation}
      \hat{\mathbf{\Psi}}^{\dagger}(\mathbf{r})=
      \left(\begin{array}{cc} \hat{\Psi}^{\dagger}(\uparrow,\mathbf{r}), & \hat{\Psi}^{\dagger}(\downarrow,\mathbf{r})
      \end{array}\right), \label{conj_FOS}
\end{equation}
where
\begin{equation}
      \hat{\Psi}(x)=\sum_{k}\psi_{k}^{S}(x)\hat{C}_{k}
\end{equation}
is the field operator with $x\equiv (\mathbf{r},\sigma)$, $\sigma \in \{ \uparrow,\downarrow \}$ and the
annihilation operator, $\hat{C}_{k}$,
for the single-electron state (SES) $\ket{\psi_k^S}$ in the central system.
The SES $\ket{\psi_k^S}$ is the eigenstate labeled by $k$
of the Hamiltonian $\hat{H}_{S}-\hbar\omega \hat{a}^{\dagger}\hat{a}$
when we set the photonic part of the vector potential $\hat{\mathbf{A}}^{\mathrm{ph}}(\mathbf{r})$
in the momentum operator,
\begin{equation}
      \hat{\mathbf{p}}(\mathbf{r})=\left(\begin{array}{c} \hat{p}_x(\mathbf{r}) \\ \hat{p}_y(\mathbf{r})
      \end{array}\right)=-i\hbar\nabla +\frac{e}{c}
      \hat{\mathbf{A}}^{\mathrm{ph}}(\mathbf{r}),   \label{mom}
\end{equation}
to zero.

The last term in \eq{H^S} indicates the quantized photon field,
where $\hat{a}^{\dagger}$ is the
photon creation operator
and $\hbar\omega$ is the photon excitation energy.
The zero point energy, which has no other implications than a constant shift of the
energy spectrum, is neglected here.
The photon field
interacts with the electron system via the vector potential
\begin{equation}
      \hat{\mathbf{A}}^{\mathrm{ph}}=A(\mathbf{e}\hat{a}+\mathbf{e}^{*}\hat{a}^{\dagger}) \label{vec_pot}
\end{equation}
with
\begin{equation}
 \mathbf{e}=\left\{  \begin{array}{cl} \mathbf{e}_x, & \mathrm{TE}_{011} \\ \mathbf{e}_y, & \mathrm{TE}_{101} \\
             \frac{1}{\sqrt{2}}\left[\mathbf{e}_x+ i\mathbf{e}_y\right], & \textrm{RH circular} \\
             \frac{1}{\sqrt{2}}\left[\mathbf{e}_x- i\mathbf{e}_y\right], & \textrm{LH circular}
\end{array} \right.
\end{equation}
for a longitudinally-polarized ($x$-polarized) photon
field ($\mathrm{TE}_{011}$), transversely-polarized ($y$-polarized)
photon field ($\mathrm{TE}_{101}$), right-hand (RH) or left-hand (LH) circularly polarized photon field.
For reasons of simplicity, we show here only the results for the $x$-polarized photon field, but comment
on the results for photon fields with other polarization.
In particular, the number of allowed transitions between MB states is much larger for circular polarization.
The electron-photon coupling
constant
\begin{equation}
      g_{\gamma}=eA a_w \Omega_w/c
\end{equation}
scales with the amplitude $A$ of
the electromagnetic field,
and the natural length scale due to the confinement of the system,
\begin{equation}
      a_w = \left(\frac{\hbar}{m^*\Omega_0}\right)^{1/2}.
\label{a_w}
\end{equation}
Our model of a photon cavity can be realized
experimentally~\cite{0034-4885-69-5-R02, 0953-4075-38-9-007, 0953-8984-20-45-454209}
by letting the photon cavity be much larger than the quantum ring
(this assumption is used in the derivation of the vector potential, \eq{vec_pot}).

\subsection{Quantum ring potential}
We model a small quantum ring with a finite width.
\begin{figure}[htbq]
       \includegraphics[width=0.45\textwidth,angle=0,bb= 88 60 446 304,clip]{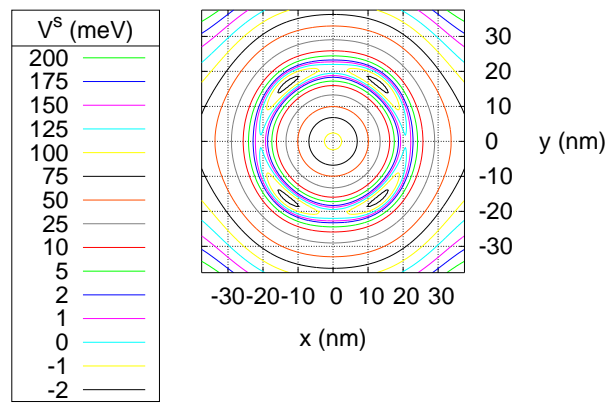}
       \caption{(Color online) Equipotential lines of the
       confinement potential $V(\mathbf{r})$ of the quantum ring.
       The equipotential lines are refined at the bottom of the ring.}
       \label{ext_pot}
\end{figure}
The quantum ring confinement potential is shown in \fig{ext_pot}.
The ring is relatively small and the electrons relatively strongly confined
to minimize the computational effort.
Mathematically, the expression for the potential is
\begin{eqnarray}
      V(\mathbf{r})&=& \sum_{i=1}^{4}V_{i}\exp
      \left[
      -\left(\beta_{xi}(x-x_{0i})\right)^2 - \left(\beta_{yi}y\right)^2
      \right] \nonumber \\
      &&+\frac{1}{2}m^* \Omega_{0}^2y^2, \label{V_S}
\end{eqnarray}
with the parameters from Tab.\ \ref{table:ringpot}
and the characteristic confinement energy in $y$-direction being
$\hbar\Omega_0 = 16.0$~meV. $x_{03}=\epsilon$ is a small numerical
symmetry breaking parameter and $|\epsilon|=10^{-5}$~nm is enough
for numerical stability.
\begin{table}
      \caption{Parameters of the ring potential $V(\mathbf{r})$.}
      \centering
      \begin{tabular}{c    |    c    |    c    |    c    |    c}
      \hline\hline
      %\cline{1-5}
      \\ [-1.4ex]
      $i$ &  $V_{i}$ in meV &  $\beta_{xi}$ in $\frac{1}{\mathrm{nm}}$
      &  $x_{0i}$ in nm &  $\beta_{yi}$ in $\frac{1}{\mathrm{nm}}$ \\ [0.5ex]
      \hline
      1 & 164.8 & 0.044 & 45 & 0 \\
      2 & 164.8 & 0.044 & -45 & 0 \\
      3 & 177.6 & 0.066 & $\epsilon$ & 0.068 \\
      6 & -80.0 & 0 & 0 & 0 \\[0.5ex]
      \hline\hline
      \end{tabular}
\label{table:ringpot}
\end{table}

\subsection{Excitation pulse and time evolution}

The time evolution is given by the Liouville-von Neumann equation
\begin{equation}
      i\hbar \frac{d}{dt}\hat{\rho}(t)=\left[ \hat{H}_{\rm MB} + \hat{W}_{\rm MB}(t), \hat{\rho}(t) \right] \label{LVN}
\end{equation}
with the MB operator $\hat{W}_{\rm MB}(t)$ representing a short excitation pulse
\begin{align}
      W(\mathbf{r},t)=&W_{d}(\mathbf{r})\exp(-\Gamma t) \nonumber \\
      &\times \sin(\omega_2 t)\sin(\omega_1 t)\theta (10\pi-\omega_2 t)
\end{align}
with the dipole potential
\begin{equation}
      W_d(\mathbf{r})=W_0 x \label{realdippot}
\end{equation}
in $x$-direction with $W_0=2.36\times 10^{-2}$ meV/nm.

In \eq{LVN}, the density operator $\hat{\rho}(t)$
and the other operators appear in the MB presentation.
Numerically, we solve the Liouville-von Neumann equation \eq{LVN}
using the time-evolution operator $\hat{U}_{\rm MB}(t)$
defined by~\cite{PhysRevB.67.161301}
\begin{equation}
      \hat{\rho}(t) = \hat{U}_{\rm MB}(t) \hat{\rho}(0) \hat{U}^{\dagger}_{\rm MB}(t)
\end{equation}
yielding the equations of motion
\begin{align}
      i\hbar \dot{\hat{U}}_{\rm MB}(t) =& \hat{H}_{\rm MB}^{\rm td}(t) \hat{U}_{\rm MB}(t),\nonumber \\
     -i\hbar \dot{\hat{U}}^{\dagger}_{\rm MB}(t) =& \hat{U}^{\dagger}_{\rm MB}(t) \hat{H}_{\rm MB}^{\rm td}(t) \label{eom}
\end{align}
with the time-dependent MB Hamiltonian
\begin{equation}
      \hat{H}_{\rm MB}^{\rm td}(t) = \hat{H}_{\rm MB} + \hat{W}_{\rm MB}(t).
\end{equation}
The time integration of \eq{eom} is done using
the Crank-Nicolson algorithm
with the initial condition $\hat{U}_{\rm MB}(0)=\mathbbm{1}$ and
with $\hat{H}_{\rm MB}^{\rm td}(0)=\hat{H}_{\rm MB}$ \cite{PhysRevB.67.161301}.

\section{Results}

For all our results, we start the system in the ground state with a photon content close to
zero and propagate the system long after the external pulse has vanished,
until the time $t=440$~ps to get a precise description of the
Fourier transform (FT) of various oscillations the system is performing.
We assume a GaAs-based
material with the electron effective mass $m^*=0.067m_e$ and background
relative dielectric constant $\kappa = 12.4$.
The single photon cavity mode has the excitation energy $\hbar \omega=6.4$~meV
and the energies of the excitation pulse $\hbar \omega_1=2.63$~meV and $\hbar \omega_2=0.658$~meV and
the decay factor $\Gamma=0.2$~ps$^{-1}$.
A magnetic field $B=10^{-5}$~T five orders of magnitude below the Aharonov-Bohm~\cite{PhysRev.115.485}
regime for our ring size is used here for the numerical purpose to lift the spin degeneracy.
The length scale (defined in \eq{a_w}) is $a_w \approx  8.431$~nm.

The electron number in the system is fixed to one electron to reduce the numerical effort and to focus the
attention on the effects of the electron-photon interaction. Therefore,
the Coulomb interaction does not have to
be considered in \eq{H^S}.

\subsection{Energetics analysis of the system}

\begin{figure}[htbq] %bb= 94 80 508 377
       \includegraphics[width=0.45\textwidth,angle=-90]{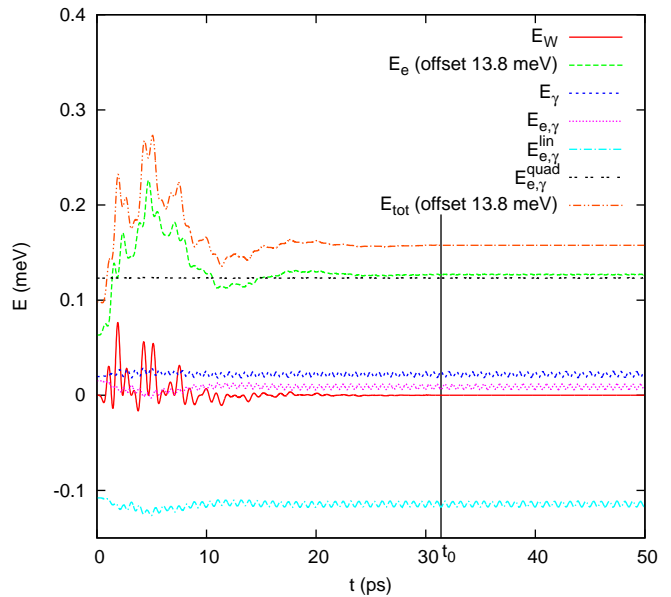}
       \caption{(Color online) Energies in the system as a function of time. The zero point energy of the photons
       is excluded. Two very positive energy curves have been reduced by a convenient offset of $13.8$~meV to smaller
       values to fall in a similar energy range as the other curves. The photon field is $x$-polarized.
       The electron-photon coupling constant $g_{\gamma}=0.5$~meV. The end of the excitation pulse is marked by a vertical line
       at $t=t_0:=10\pi/\omega_2=31.4$~ps. Energies are defined by Equations (\ref{Ee})-(\ref{eg}).}
       \label{energy}
\end{figure}

The quantum ring system is filled with an electron coupled to cavity photons of a single
frequency and is influenced by a weak dipole excitation pulse in the $x$-direction. To understand the time-dependency
of the different components of the
system, we take first a look at their energy contents in \fig{energy}.
The total energy (without zero point energy of the photons)
\begin{equation}
 E_{\rm tot}(t)={\rm Tr} [\hat{\rho}(t) \hat{H}_{\rm MB}^{\rm td}(t)]
\label{Ee}
\end{equation}
is a constant after the excitation pulse ($t>t_0:=10\pi/\omega_2=31.4$~ps),
as the system can then be considered
to be closed relative to its environment.
However, the MB system energy can be redistributed and oscillate between the electron
and the photons after the excitation by the pulse. More correctly,
there are three energy contributions after the pulse, the energy of the electron
\begin{equation}
      E_{e}(t)={\rm Tr} [\hat{\rho}(t) \hat{H}_{\rm MB}^{e}]
\end{equation}
with
\begin{equation}
      \hat{H}_{\rm MB}^{e}=\int d^2 r\; \hat{\mathbf{\Psi}}^{\dagger}(\mathbf{r})
      \left[\frac{\hat{\mathbf{p}}'^2}{2m^{*}} +V(\mathbf{r})\right]\hat{\mathbf{\Psi}}(\mathbf{r})
\end{equation}
and
\begin{equation}
      \hat{\mathbf{p}}'=-i\hbar\nabla,
\end{equation}
the energy of the photons (without zero point energy)
\begin{equation}
      E_{\gamma}(t)={\rm Tr} [\hat{\rho}(t)\hbar\omega \hat{a}^{\dagger}\hat{a}] \label{phen}
\end{equation}
and the energy due to the interaction term of the electron and the photons
\begin{equation}
 E_{e,\gamma}(t)={\rm Tr} [\hat{\rho}(t) \hat{H}_{\rm MB}^{e,\gamma}]
\end{equation}
with
\begin{equation}
 \hat{H}_{\rm MB}^{e,\gamma}=\hat{H}_{\rm MB, lin}^{e,\gamma}+\hat{H}_{\rm MB, par}^{e,\gamma},
\end{equation}
\begin{equation}
      \hat{H}_{\rm MB, lin}^{e,\gamma}=\int d^2 r\; \hat{\mathbf{\Psi}}^{\dagger}(\mathbf{r})
      \left[\frac{\hat{\mathbf{p}}'\hat{\mathbf{A}}^{\rm ph}(\mathbf{r})+\hat{\mathbf{A}}^{\rm ph}
      (\mathbf{r})\hat{\mathbf{p}}'}{2m^{*}c/e}\right]\hat{\mathbf{\Psi}}(\mathbf{r})
\end{equation}
and
\begin{equation}
     \hat{H}_{\rm MB, par}^{e,\gamma}=\int d^2 r\; \hat{\mathbf{\Psi}}^{\dagger}(\mathbf{r})
     \left[\frac{e^2\left(\hat{\mathbf{A}}^{\rm ph}(\mathbf{r})\right)^2}{2m^{*}c^2}\right]\hat{\mathbf{\Psi}}(\mathbf{r}).
\label{eg}
\end{equation}
The electron energy is by far the largest contribution (around $13.9$~meV), the photon energy
is about $3.2$~meV when adding the zero point energy and the electron-photon interaction energy
is much smaller than $0.1$~meV. However, when separating the interaction energy into two terms with a linear,
\begin{equation}
       E_{e,\gamma}^{\rm lin}(t)={\rm Tr} [\hat{\rho}(t) \hat{H}_{\rm MB, lin}^{e,\gamma}], \label{linA}
\end{equation}
or quadratic,
\begin{equation}
      E_{e,\gamma}^{\rm quad}(t)={\rm Tr} [\hat{\rho}(t) \hat{H}_{\rm MB, par}^{e,\gamma}],
\end{equation}
dependency on the vector potential $\hat{\mathbf{A}}^{\mathrm{ph}}(\mathbf{r})$, it can be seen that each of
this contribution is larger than $0.1$~meV,
but with opposite signs.

All energy contributions show similar oscillations after the excitation pulse and pass energy
between each other. In particular, the oscillation of the linear interaction energy, \eq{linA},
and photon energy, \eq{phen}, are strong and almost in anti-phase.
In the following, we will concentrate on the oscillations of the photon energy or more precisely said,
the oscillations in the mean photon number, which is proportional to the photon energy.

During the excitation pulse, there is an additional energy appearing due to the excitation pulse itself
\begin{equation}
      E_{W}(t)={\rm Tr} [\hat{\rho}(t) \hat{W}_{\rm MB}(t)].
\end{equation}
The excitation pulse can also increase or decrease the total energy of the
otherwise closed system. The change of the total energy is relatively small
compared to the photon excitation energy $\hbar \omega$.
In this sense, the excitation pulse can be considered to be weak
meaning that it can not change the mean photon number by a whole photon.

\subsection{Dipole moment oscillations of the charge density}

\begin{figure}[htbq] %bb= 94 80 508 377
       \includegraphics[width=0.35\textwidth,angle=-90]{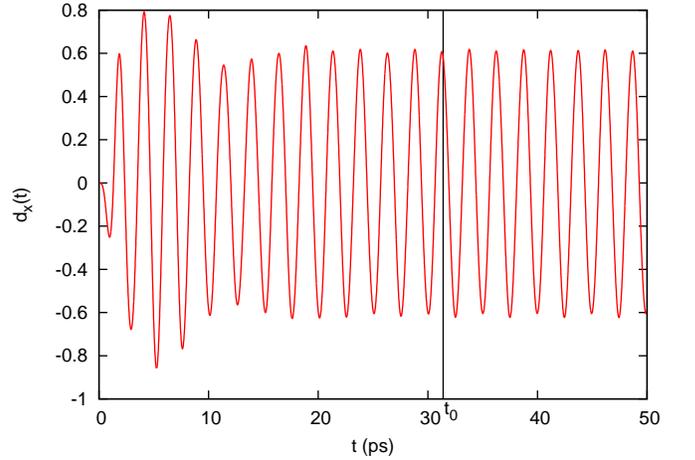}
       \caption{(Color online) Dipole moment oscillations  of the charge density distribution in
       $x$-direction as a function of time. The photon field is $x$-polarized. The electron-photon
       coupling constant $g_{\gamma}=0.5$~meV. The end of the excitation pulse is marked by a
       vertical line at $t=t_0$.}
       \label{dipole}
\end{figure}

It is interesting to investigate the response of the charge density to the external excitation of
the quantum ring system.
Excited by the dipole excitation pulse in $x$-direction,
the center of charge (dipole moment of the charge density)
\begin{equation}
      \mathbf{d}(t)={\rm Tr}[\hat{\rho}(t) \hat{\mathbf{d}}]
\end{equation}
with $\hat{\mathbf{d}}$ being the dipole operator,
\begin{equation}
      \hat{d}_{i}=e \int d^2 r\; r_{i} \hat{\mathbf{\Psi}}^{\dagger}(\mathbf{r}) \hat{\mathbf{\Psi}}(\mathbf{r}),
\end{equation}
where $i=x,y$, oscillates as is shown in \fig{dipole}. The oscillations are a superposition of periodic oscillations
after the excitation pulse, but not during the time, the pulse is exciting the system. We therefore exclude
the excitation time interval
$t<t_0$ for a further analysis of the oscillations.

\begin{figure}[htbq] %bb= 94 80 508 377
       \includegraphics[width=0.35\textwidth,angle=-90]{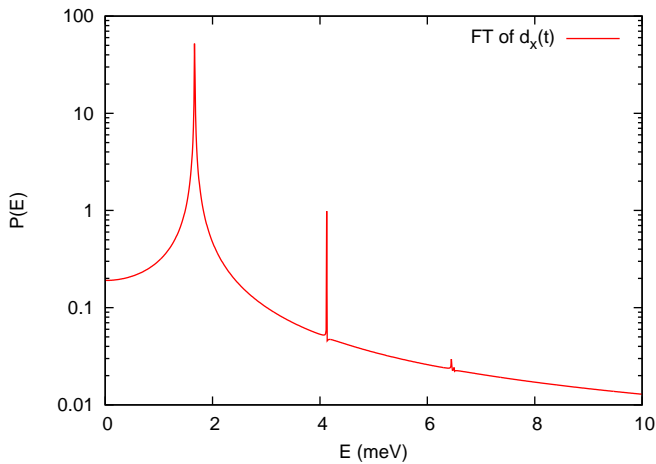}
       \caption{(Color online) Logarithmic plot of the FT of the dipole moment oscillations of the
       charge density distribution as a function of energy after the excitation pulse ($t>t_0$).
       The photon field is $x$-polarized.
       The electron-photon coupling constant $g_{\gamma}=0.5$~meV.}
       \label{fft}
\end{figure}

Figure \ref{fft} shows the FT of the center of charge oscillations. It is mainly composed of two peaks,
a strong one at $E=1.66$~meV and a much
weaker one at $E=4.13$~meV. (Notice the logarithmic scale of the peak height.)
For $x$- or $y$-polarized (linearly polarized)
cavity photon field the center of charge oscillates in $x$-direction, meaning that $d_y(t)$ is vanishing.
We have seen, however, that a circularly polarized cavity photon field leads to center of charge oscillations
in both $d_x(t)$ and $d_y(t)$, even
though the excitation pulse is only a dipole excitation in $x$-direction. Alternatively, a strong magnetic field
leads also to center of charge oscillations in both the $x$- and $y$-direction, even though the cavity photon field is
linearly polarized.

\subsection{Fourier analysis of the mean photon number oscillations}

\begin{figure}[htbq] %bb= 94 80 508 377
       \includegraphics[width=0.35\textwidth,angle=-90]{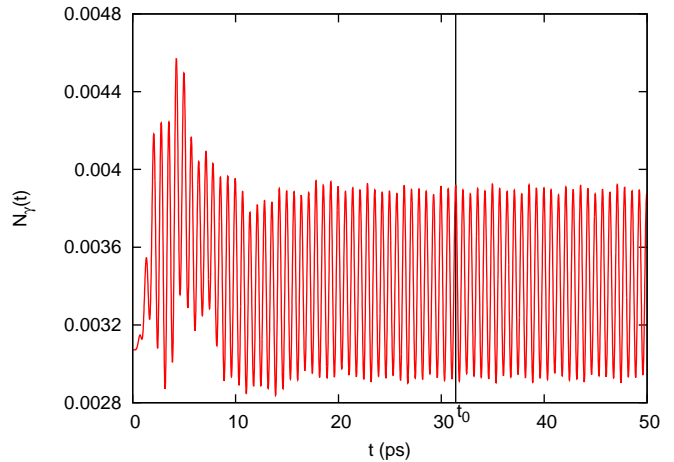}
       \caption{(Color online) Mean photon number oscillations as a function of time.
       The photon field is $x$-polarized. The electron-photon coupling constant $g_{\gamma}=0.5$~meV.
       The end of the excitation pulse is marked by a vertical line at $t=t_0$.}
       \label{mean_ph}
\end{figure}

We have seen earlier that the photon energy oscillates also in time.
In \fig{mean_ph}, we show the related mean photon number
\begin{equation}
      \langle N_{\gamma}(t) \rangle ={\rm Tr} [\hat{\rho}(t) \hat{a}^{\dagger}\hat{a}]
\end{equation}
as a function of time, which describes the number of cavity photons in the system.
Similar to \fig{dipole}, the oscillations are a superposition of periodic oscillations
after the excitation pulse, but the main components are of a higher frequency.

\begin{figure}[htbq] %bb= 94 80 508 377
       \includegraphics[width=0.35\textwidth,angle=-90]{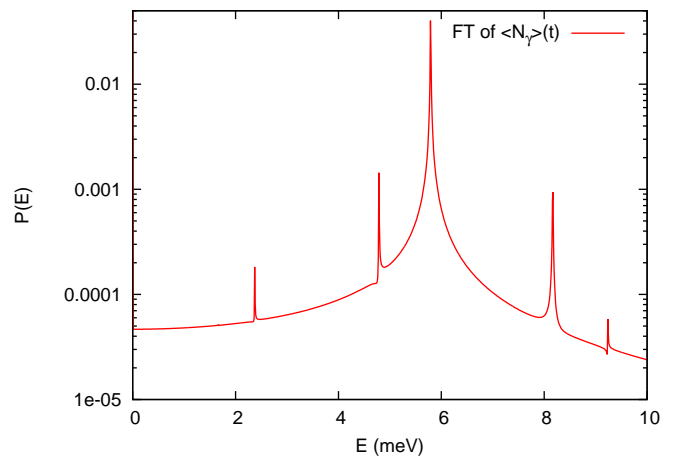}
       \caption{(Color online) Logarithmic plot of the FT of the mean photon number
       oscillations as a function of energy after the excitation pulse ($t>t_0$). The photon field is $x$-polarized.
       The electron-photon coupling constant $g_{\gamma}=0.5$~meV.}
       \label{fft_mean}
\end{figure}

This can be seen more clearly from the Fourier analysis in \fig{fft_mean}, which shows the main peaks at
the energies $E=2.37$~meV, $E=4.79$~meV, $E=5.79$~meV and $E=8.17$~meV, which would all be visible
in a linear plot with the largest Fourier peak
at the relatively high energy $E=5.79$~meV. As mentioned earlier the oscillations of the electron energy and
electron-photon interaction energy are composed of a similar frequency spectrum as the mean photon number.
But why are the center of charge oscillations with a different spectral composition (\fig{fft})?

\subsection{Comparison with the many-body spectrum}

\begin{figure}[htbq] %bb= 94 80 508 377
       \includegraphics[width=0.45\textwidth,angle=-90]{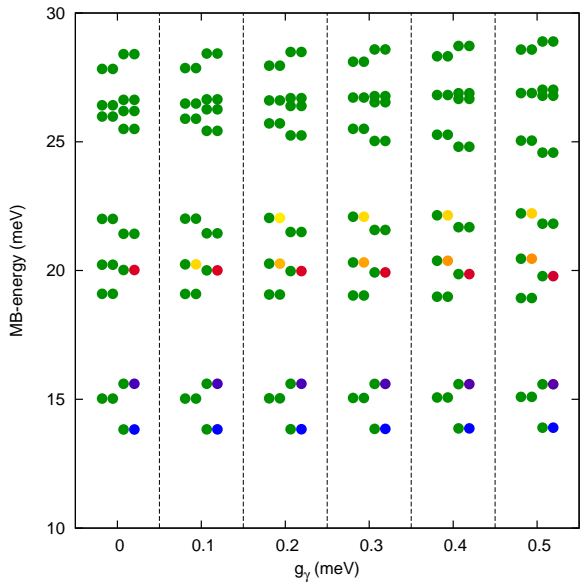}
       \caption{(Color online) MB energy spectrum of the system Hamiltonian \eq{H^S}
       versus the electron-photon coupling constant $g_{\gamma}$ for $x$-polarized photon field.
       The occupation of the states at $t=440$~ps is indicated by the color of the dots with the continuous
       color spectrum from yellos over red to blue corresponding to the range of the occupation 
       number $[5\times 10^{-9},1]$.
       If the occupation number is below $5\times 10^{-9}$, the dots are colored green.
       The occupation of all MB states with an energy above $30$~meV is below $5\times 10^{-9}$.
       Due to the small energy differences between MB states it was necessary to delocate the dots
       slightly along the $g_{\gamma}$-axis such that their occupation can be clearly recognized. This is
       not indicating slightly different $g_{\gamma}$-values, all dots belong exclusively
       to $g_{\gamma}=0,0.1,0.2,0.3,0.4,0.5$.}
       \label{MB_spec_vg_x_B000001_npi10_lowercut}
\end{figure}

To understand the Fourier components of the oscillations in the center of charge and the mean photon number,
we have a look at the MB energy spectrum of the quantum ring system and the occupation of the MB states at $t=440$~ps
in \fig{MB_spec_vg_x_B000001_npi10_lowercut}. Only a few states are with an occupation above $5\times 10^{-9}$,
which we find all relevant to understand the visible peaks in a linear plot.
(The sum over the occupation of all MB states is one.)
This is showing that the dipole excitation pulse is not very strong, it is however strong enough that we would
expect differences in the results, when we would compare to a linear
response calculation. Furthermore, the relevant MB states lie all at rather low energies, hinting at the fact that the
selected $N_{\rm MB}=120$ MB states are sufficient to predict the time-evolution of the system with reasonable
accuracy (only the lowest MB states are shown in \fig{MB_spec_vg_x_B000001_npi10_lowercut}).
The energetic position and occupation of the MB states changes slightly with the electron-photon
coupling constant $g_{\gamma}$.

\begin{figure}[htbq] %bb= 94 80 508 377
       \includegraphics[width=0.49\textwidth,angle=-90]{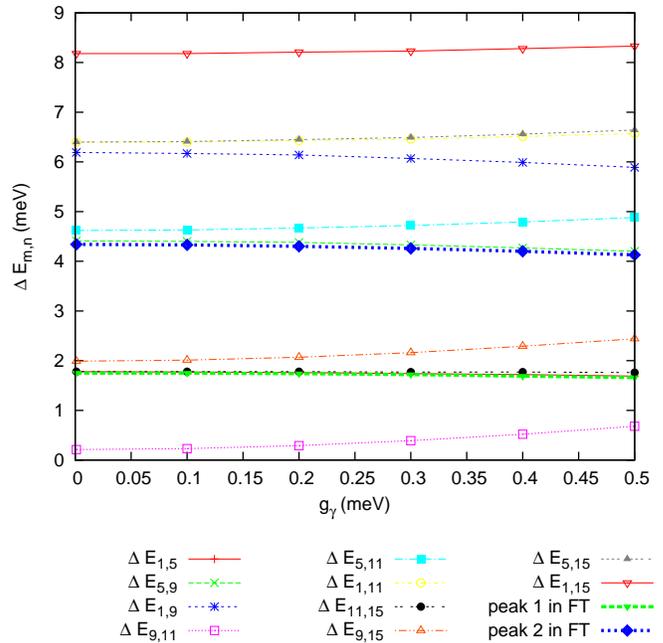}
       \caption{(Color online) Bohr energies $\Delta E_{m,n}$ between the MB states $m$ and $n$ of the system
       Hamiltonian \eq{H^S} and energetic location of the main peaks of the FT
       of the dipole moment of the charge density distribution versus the electron-photon coupling constant $g_{\gamma}$.
       The photon field is $x$-polarized.}
       \label{levels_x_dipole}
\end{figure}

To compare the MB spectrum \fig{MB_spec_vg_x_B000001_npi10_lowercut} to the Fourier analysis of the center of charge
oscillations, we look at \fig{levels_x_dipole}.
Here, we have plotted the Bohr energies $\Delta E_{m,n}$ between all MB states with occupation above $5\times 10^{-9}$
and compare them to the peaks of the FT of the center of charge oscillations.
We can see that not all transitions are likely or allowed and that some selection rules apply.
In particular, we find the allowed transitions by
considering the value of the matrix element
\begin{equation}
      W_{m,n}=\bra{m}\hat{W}_{d,{\rm MB}}\ket{n}
\end{equation}
with $\hat{W}_{d,{\rm MB}}$ being the MB representation of the dipole potential \eq{realdippot}.
The Bohr energies of the allowed transitions with large $|W_{m,n}|$ are $\Delta E_{1,5}$, $\Delta E_{5,9}$
and $\Delta E_{11,15}$.
These transitions are allowed for the whole $g_{\gamma}$-range. In addition, for $g_{\gamma}>0$,
the Bohr energies $\Delta E_{9,11}$, $\Delta E_{1,11}$ and $\Delta E_{5,15}$ are allowed.
The Bohr energies $\Delta E_{1,5}$ and $\Delta E_{5,9}$ can be recognized as peaks of the FT of the center
of charge oscillations. The occupation of the higher MB states $n>9$ is very low, such that the transition
$\Delta E_{11,15}$ is very weak. The same argument applies for the transitions, which are only allowed for
$g_{\gamma}>0$. When comparing the two allowed and likely FT peaks in their strength, the peak with
$\Delta E_{1,5}$ is dominant over the peak with $\Delta E_{5,9}$. This can be directly correlated to the fact
that the occupation of the MB state $n=1$ is much larger
than the occupation of the MB state $n=9$.

\begin{figure}[htbq] %bb= 94 80 508 377
       \includegraphics[width=0.49\textwidth,angle=-90]{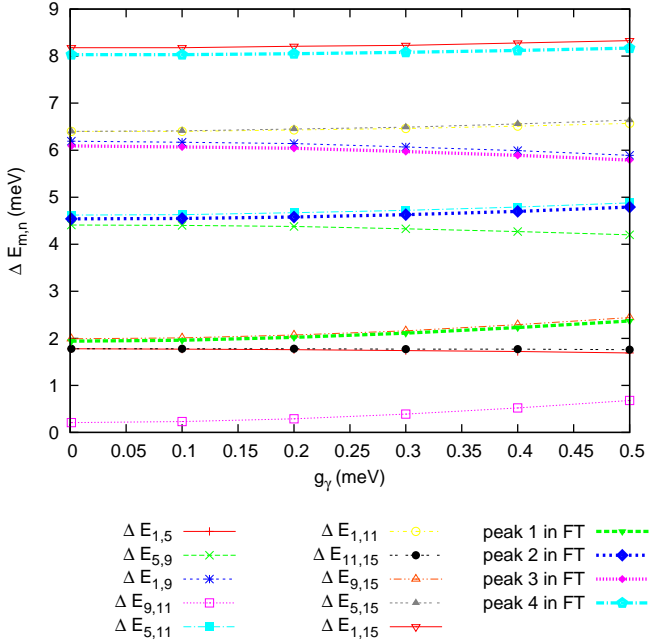}
       \caption{(Color online) Bohr energies $\Delta E_{m,n}$ between the MB states $m$ and $n$ of
       the system Hamiltonian \eq{H^S} and energetic location of the main peaks of the FT
       of the mean photon number versus the electron-photon coupling constant $g_{\gamma}$.
       The photon field is $x$-polarized.}
       \label{levels_x_photon}
\end{figure}

Figure \ref{levels_x_photon} shows the comparison of the MB spectrum \fig{MB_spec_vg_x_B000001_npi10_lowercut}
to the Fourier analysis of the mean photon number oscillations.
We note here, that we can associate a fractional photon content $\mu$ to each MB state.
The deviation of the photon content from integer numbers increases in general with the electron-photon
coupling strength $g_{\gamma}$. Still, we can approximately state that the MB states $n=1,5,9$ are with
a photon content close to zero and $n=11,15$ are with a photon content close to one. The transitions observed
as FT peaks of the mean photon number are $\Delta E_{9,15}$, $\Delta E_{5,11}$, $\Delta E_{1,9}$ and
$\Delta E_{1,15}$. Except for the strongest peak with $\Delta E_{1,9}$, the photon content is changing in all other
transitions by approximately one ($\Delta \nu \approx 1$). For the peak with $\Delta E_{1,9}$,
the photon content difference of the states
is $\Delta \nu = 0.096$ for $g_{\gamma}=0.5$~meV and gets smaller with decreasing $g_{\gamma}$.
However, it is clear from \fig{MB_spec_vg_x_B000001_npi10_lowercut} that the occupation of the
excited MB state for the transitions with $\Delta E_{9,15}$, $\Delta E_{5,11}$ and $\Delta E_{1,15}$
becomes also smaller with decreasing $g_{\gamma}$. As a consequence,
we found all Fourier peaks of the mean photon number to become smaller with decreasing $g_{\gamma}$,
but their relative strength is almost conserved
(i.e.\ the main change of \fig{fft_mean} with $g_{\gamma}$ is only the scaling of the $y$-axis).
Therefore, we have to consider both the change in the photon content $\Delta \nu$
and the occupations of the two MB states associated with the transition
to be able to say something about the selection rules governing the optical transitions.
As a side remark, we could also associate an angular momentum $M$ to each MB state with increasing deviations from integer
angular momentum numbers $m=M/\hbar$ with increasing electron-photon coupling strength $g_{\gamma}$.
The absolute value of the angular momentum number $|m|\approx 0$ for the MB states $n=1,11$; $|m|\approx 1$
for the MB states $n=5,15$; and $|m|\approx 2$ for the MB state $n=9$.
The MB states corresponding to the FT peaks of the mean photon number show a difference of $\Delta |m|\approx 1$ with
one exception with the Bohr energy $\Delta E_{1,9}$, where $\Delta |m|\approx 2$.

\begin{figure}[htbq] %bb= 94 80 508 377
       \includegraphics[width=0.58\textwidth,angle=-90]{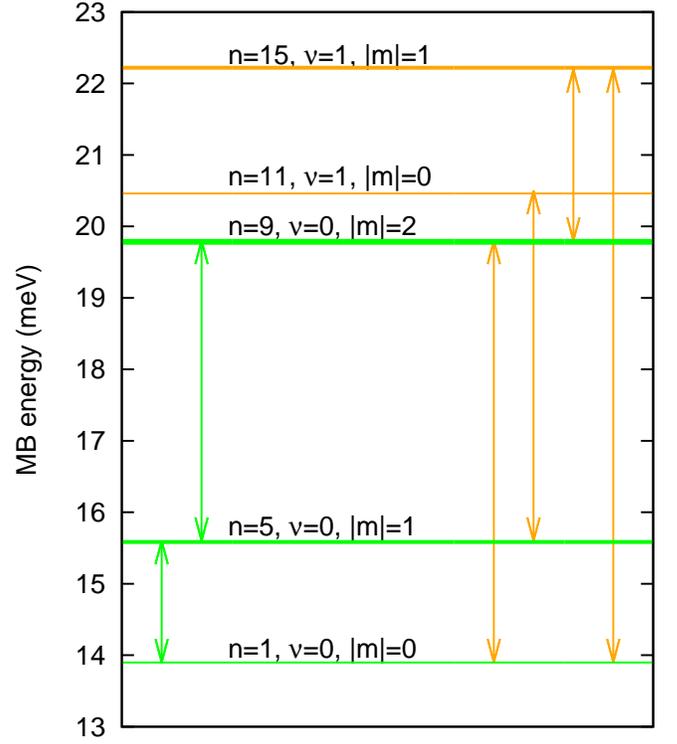}
       \caption{(Color online) Mostly occupied MB energy levels and transitions between them.
       The number $n$ is the level number for energetic ordering of the MB states, $\nu$ is the approximate
       photon content and $|m|$ the approximate absolute value of the angular momentum number.
       The line width increases with $|m|$ in the cartoon and the green color means an approximate photon
       content $\nu \approx 0$, while the orange color means an approximate photon content $\nu \approx 1$.
       The green transitions correspond to the Bohr
       energies seen as main peaks in the FT of the center of charge oscillations and the orange transitions correspond to
       the Bohr energies seen as main peaks in the FT of the mean photon number oscillations.
       The electron-photon coupling constant $g_{\gamma}=0.5$~meV. The photon field is $x$-polarized.}
       \label{transitions}
\end{figure}

The approximate values of the absolute value of the
angular momentum numbers $m$ and photon contents $\nu$ for the
mostly populated MB levels are depicted in \fig{transitions}. It
shows also the Bohr energies, which describe the frequencies of the
center of charge oscillations and the frequencies of the
oscillations of the mean photon number. One could pose the
interesting question concerning the transition with Bohr energy
$\Delta E_{1,9}$ of the latter type, where $\Delta |m|\approx 2$ and
$\Delta \nu \approx 0$, whether this transition would in fact be
composed of two processes, each with $\Delta |m|\approx 1$ and
possibly a photon content difference $\Delta \nu \approx 1$ in the
first and $\Delta \nu \approx -1$ in the second process.

\section{Conclusions}

We have studied the non-linear response of a quantum ring system
to a short dipole excitation pulse.
The quantum ring system is coupled to a photon cavity using exact numerical diagonalization.
We have seen that the many-level description is essential to describe properly the physical response of the system.
The short pulse excites oscillations of the energy between the electron and cavity photons.
The coupling energy between the photons and electron is small as the
linear and quadratic term in the vector potential are of opposite signs.
We find center of charge oscillations in the direction of the dipole excitation of the pulse.
The direction of the linear polarized cavity photon field does not influence the center of charge oscillation
direction. A circularly polarized cavity photon field or magnetic field, however, changes the direction.
The oscillations of the mean photon number and center of charge have a different Fourier spectrum,
but are all reflected in transition energies between MB levels.
The oscillator strengths for the center of charge oscillation spectrum are given by selection rules
due to the matrix elements of the dipole potential with the MB states corresponding to a MB transition
(i.e.\ the geometrical symmetry of the MB states) and the population of the levels.
For the selection rules governing the mean photon number oscillations, the difference of the photon content
of the MB states determining a transition
and the MB level population play a crucial role.

In summary, we have supplied a rather small amount
of energy to the system with an excitation pulse with a rather broad
frequency spectrum, we are thus not exciting isolated resonances in
the system, but instead probing its response to
a broad frequency range.  It is demonstrated why
both the linear and quadratic electron-photon interactions are
necessary together with a large section of the system states.

\section*{Acknowledgments}
      This work was financially supported by the Icelandic Research
      and Instruments Funds, the Research Fund of the University of Iceland, and
      the Ministry of Science and Technology,
       Taiwan through Contract No.\ MOST 103-2112-M-239-001-MY3. We acknowledge also support
      from the computational facilities of the Nordic High Performance Computing (NHPC).

\twocolumngrid
%
%---------------------------------------------
%
%
\bibliographystyle{apsrev4-1}
\end{document}